\documentclass[%
aps,prb,superscriptaddress,twocolumn,floatfix,10pt,longbibliography]{revtex4-2}
\usepackage[utf8]{inputenc}
\usepackage{amsmath,amssymb}
\usepackage{empheq}
\usepackage{float}
\usepackage{lipsum}
\usepackage{slashed}
\usepackage{mathtools,cuted}
\usepackage{esvect}
\usepackage[none]{hyphenat}
\usepackage{multirow}
\usepackage{xcolor}
\usepackage{soul}
\usepackage[export]{adjustbox}
\usepackage{graphicx}
\usepackage{dcolumn}
\usepackage{bm}
\usepackage{hyperref}
\usepackage{physics}
\usepackage[mathlines]{lineno}
\usepackage[capitalise]{cleveref}
\usepackage{bbm}
\usepackage{orcidlink} 

\begin{document}
	
\title{Topological phases of the interacting SSH model: an analytical study}

\author{E. Di Salvo}
    \affiliation{Institute for Theoretical Physics, Utrecht University, 3584CC Utrecht, The Netherlands,}%
\author{A. Moustaj}
    \affiliation{Institute for Theoretical Physics, Utrecht University, 3584CC Utrecht, The Netherlands,}
\author{C. Xu}
    \affiliation{Department of Physics and Materials Science, University of Luxembourg, L-1511 Luxembourg, Luxembourg,}
    \affiliation{Institute of Theoretical Physics, Technische Universität Dresden, 01069 Dresden, Germany,}
\author{L. Fritz}
    \affiliation{Institute for Theoretical Physics, Utrecht University, 3584CC Utrecht, The Netherlands,}
\author{A.~K. Mitchell}
    \affiliation{School of Physics, University College Dublin, Belfield, Dublin 4, Ireland,}
    \affiliation{Centre for Quantum Engineering, Science, and Technology, University College Dublin, Ireland.}
\author{C. Morais Smith}
    \affiliation{Institute for Theoretical Physics, Utrecht University, 3584CC Utrecht, The Netherlands,}
\author{D. Schuricht}
    \affiliation{Institute for Theoretical Physics, Utrecht University, 3584CC Utrecht, The Netherlands,}

	\date{\today}
	
\begin{abstract}
The interacting SSH model provides an ideal ground to study the interplay between topologically insulating phases and electron-electron interactions. We study the  polarization density as a topological invariant and provide an analytic treatment of its behavior in the low-energy sector of the one-dimensional interacting SSH model. By formulating the topological invariant in terms of Green's functions, we use the low-energy field theory of the Thirring model to derive the behavior of the polarization density. We show that polarization density in the continuum theory describes the usual topological insulating phases. Still, it contains an extra factor from the fields' scaling dimensions in the low-energy quantum field theory. We interpret this as a measure of the modified charge of the new excitations in the system. We find two distinct contributions: a renormalization of the electronic charge $e$ of a Fermi liquid because of quasiparticle smearing and an additional contribution coming from the topological charge of the soliton arising in the bosonized version of the Thirring model, the sine-Gordon model. 
\end{abstract}
	
\maketitle{}
	
	
\section{Introduction}\label{sec:Intro}

In the past two decades, noninteracting topological phases, traditionally analyzed through electronic band theory, have garnered significant attention due to the discovery of materials exhibiting unconventional electronic properties \cite{Qi2011TopologicalSuperconductors,Chiu2016ClassificationSymmetries}.
Most notably, the prediction of the quantum spin Hall effect in graphene by Kane and Mele \cite{Kane2005QuantumGraphene,Kane2005Z2Effect} initiated intense research in the field of topological insulators, which are materials with insulating bulk and conducting boundaries.
Since then, the theoretical understanding of these systems reached a high level of sophistication.
For example, the bulk-boundary correspondence for two-dimensional (2D) time-reversal symmetry broken insulators states that whenever the bulk Hall resistivity of the system is quantized, there exist gap traversing states, which are located at the boundary of the system \cite{Dolcetto2015EdgeInsulators}.
On the other hand, a 2D topological insulator respecting time-reversal symmetry will exhibit the quantum spin Hall effect, where the boundary states carry spin currents instead of charge currents.
This type of behavior does not only exist in 2D but also in 3D and in 1D, albeit when subjected to different spectral and spatial symmetry constraints \cite{Bradlyn2017TopologicalChemistry}.
In 3D, for example, a strong topological insulator respects time-reversal symmetry.
It has conducting surface states, which are very robust due to spin-momentum locking and cannot be destroyed by the presence of disorder or any symmetry-respecting perturbation \cite{Fu2007TopologicalDimensions,Hsieh2009ARegime}.
It also features a rich variety of additional possible effects, such as quantized magnetoelectric polarizability, providing a realization of axion electrodynamics in a condensed-matter setting \cite{Essin2009MagnetoelectricInsulators,Wilczek1987TwoElectrodynamics}.
In 1D, systems possessing inversion symmetry feature a quantized polarization density.
This fact can be attributed to two significant results: the formulation of the polarization as a Berry phase \cite{King-Smith1993TheorySolids}, and the fact that the Berry phase is quantized to $\phi=0,\pi$ for inversion symmetric systems \cite{Zak1989BerrysSolids}.
In this case, it is possible to have anomalous polarization corresponding to fractionalized charges on the two boundaries of the system.
The overarching theme in all of these cases is that one can compute bulk quantities of these systems and relate them to a corresponding boundary theory.

Additionally, they are called topological insulators because these bulk quantities are often expressed as topological invariants \cite{Ludwig2016TopologicalBeyond,Chiu2016ClassificationSymmetries}.
For example, if an inversion-symmetric 1D system also possesses sublattice symmetry, it is possible to define a winding number $\nu$, directly proportional to the quantized Berry phase, $\phi=\pi\nu$.
Another example would be a time-reversal symmetry broken 2D system, in which the Hall conductivity is quantized and given by a Chern number \cite{Thouless1982QuantizedPotential}.
This is why physical properties related to these topological invariants are very robust, as long as one respects the symmetries of the system and its gap is preserved. 

All these systems may be described by a theoretical framework based on noninteracting electrons \cite{Ludwig2016TopologicalBeyond,Chiu2016ClassificationSymmetries}. This is usually justified because many materials can be described by Fermi liquid theories in 2D and 3D at low energies. However, this description can break down in strongly correlated systems. Furthermore, in 1D, no such description is possible, and one cannot ignore the effects of electron-electron interactions. This means that the most natural state in 1D is a strongly correlated non-Fermi liquid, an example of which is the Tomonaga-Luttinger liquid \cite{Haldane1981LuttingerGas,Voit1995One-dimensionalLiquids,Giamarchi2003QuantumDimension}. 

Most of the time, topological invariants are not connected to observables in a simple way. A notable example is the Chern number and its relation to the transverse Hall conductivity. This has the main advantage that one can use standard many-body approaches in a meaningful way. In 1D, it appears that in the non-interacting case, the polarization density and the topological invariant can have the same form. While it is unclear how and when the topological invariants can be extended to a strongly interacting 1D system, this can be done with the polarization density. We find that it serves as a marker of topology and, in the existence of edge modes, also in the case of a 1D interacting system, which is not easily connected to non-interacting fermions.

The effect of interactions in topological insulators is not a new subject, and a significant number of studies have already been conducted in this direction \cite{Rachel2018InteractingReview,Gurarie2011Single-particleInsulators,Manmana2012TopologicalSystems,Chen2018WeaklyApproach,Sirker2014BoundaryModel,Marques2017MultiholeInteractions,Yahyavi2018VariationalModel,Jin2023BosonizationModel,Melo2023TopologicalModel}. For instance, many topological invariants have been generalized to expressions that hold in the many-body case, or in weakly interacting topological insulators \cite{Gurarie2011Single-particleInsulators,Manmana2012TopologicalSystems,Chen2018WeaklyApproach}.
In 1D, the paradigmatic model for a topological insulator studied extensively is the SSH chain \cite{Su1979SolitonsPolyacetylene}.
It is a tight-binding model of noninteracting fermions with nearest-neighbor alternating hopping strengths. Depending on the sign of the difference between the two hopping parameters, both a trivial phase and a topological phase can arise.
In the topological phase, two degenerate edge modes exist at $E=0$ due to sublattice symmetry.
The inclusion of interactions has been explored in various works \cite{Sirker2014BoundaryModel,Marques2017MultiholeInteractions,Yahyavi2018VariationalModel,Jin2023BosonizationModel,Matveeva2024WeaklyApproach,Chen2018WeaklyApproach,Wagner2023MottZeros,Mondal2021TopologicalDimerization,Huang2024Interaction-inducedTemperature}, relying on numerical or perturbative approaches.
Some other works have explored extensions of the model to longer range hoppings \cite{Zhou2023ExploringModel} or to ladder systems \cite{Nersesyan2020PhasePoint}.
The general result is that the topological phase usually survives the presence of interactions up to some threshold value of the interaction strength, after which the system transitions to a charge-density wave state. Another noteworthy study of the SSH model was conducted in Ref.~\cite{deLeseleuc2019ObservationAtoms}, where they experimentally realized a many-body topological phase of interacting hard-core bosons. Unlike their fermionic counterparts, the many-body bosonic representation of the protecting symmetry permits next-nearest neighbor couplings, which break sublattice symmetry in the single-particle picture. Remarkably, this implies that the many-body topological phase remains intact despite the addition of such perturbations.


In this work, we explore the polarization density as a topological invariant and provide an analytic treatment of its behavior in the low-energy sector of the interacting fermionic SSH model, with symmetry-preserving interactions. Studying the polarization density under interactions is meaningful because it is a physical observable. We achieve this by formulating the polarization density in terms of Green's functions, which are suitable for a many-body approach in the interacting case. This method is similar to the quantum Hall effect, where the transverse conductivity is quantized and proportional to the filling fraction, identifiable as a topological invariant. This identification justifies extending the study of topological invariants to interacting cases as long as the related physical observable is well-defined and measurable. Using a field theory formalism at low energy leverages many exact results in 1D for integrable models, in contrast to higher-dimensional cases.
A similar study was done for the non-Hermitian SSH model with random disorder \cite{Moustaj2022FieldModels}.
By combining various arguments, we derive an exact expression of the polarization density and provide a physical interpretation of the results, consistent with previous studies.
This paper's main result is that the continuum theory's polarization density describes the same two topologically distinct insulating phases as for the non-interacting case.
Still, it now contains an extra factor from the fields' scaling dimensions in the low-energy quantum field theory.
The interpretation is that this measures the altered nature of the excitations in the system. There are two contributions: the renormalization of the electronic charge due to quasiparticle smearing and an additional contribution from the soliton's topological charge. The latter cannot be explained within the Fermi liquid framework and highlights the difference between the effects of interactions in 1D systems compared to higher dimensions \cite{Giamarchi2003QuantumDimension}. \\

This paper is structured as follows. In \cref{Sec: Lattice SSH model}, we describe the noninteracting SSH model, its topological invariant, and the type of interactions that we will be considering. In \cref{Sec: Response Ext Field}, we formulate the polarization density as a response to an external electric field. In \cref{Sec: Polarinteraction}, we explore the effect of interactions on the polarization density. We also provide an intuitive understanding of the result, stating that an adiabatic charge pumping procedure from a trivial to a topological phase would pump one quantum of the renormalized charge, which we have calculated exactly. Finally, in \cref{Sec: Conclusion}, we present our concluding remarks and give an outlook to additional extensions of our results beyond the regimes considered in this work.

\section{Lattice SSH model}\label{Sec: Lattice SSH model}
\begin{figure}[!hbt]
    \centering
    \includegraphics[width=\columnwidth]{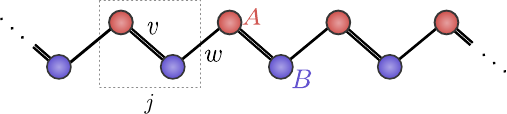}
    \caption{The SSH chain, with sublattice A (top) and B (bottom) colored in red and blue, respectively. The dotted box shows a choice of unit cell. The intracell hopping is $v$ and the intercell hopping is $w$.}
    \label{fig: SSH chain}
\end{figure}
The SSH model was originally introduced to describe the behavior of electrons and the emergence of solitonic defects in a polyacetylene molecule \cite{Su1979SolitonsPolyacetylene}.
It is a nearest-neighbor tight-binding model for spinless fermions hopping on a 1D chain with two alternating bond strengths (see \cref{fig: SSH chain}).
The Hamiltonian of the SSH model is given by
\begin{equation*}
    \begin{split}
        \hat{\mathcal{H}}&=v\sum_{j=1}^N c^\dagger_{A,j}c_{B,j}+w\sum_{j=1}^N c^\dagger_{B,j}c_{A,j+1}+\text{h.c.}, \\
        &=\int_{-\pi/a}^{\pi/a}\frac{\dd k}{2\pi}\begin{pmatrix}c^\dagger_{A,k}, & c^\dagger_{B,k}\end{pmatrix}H(k)\begin{pmatrix}c_{A,k} \\ c_{B,k}\end{pmatrix},
    \end{split}
\end{equation*}
where $v$ and $w$ are intracell and intercell hopping, respectively. 
The sum in the first line runs over cell indices, and $A,B$ refers to the two sites inside a unit cell with a lattice constant $a$ and periodic boundary conditions at the ends of the chain. 
The momentum space formulation is shown in the second line, where we have taken the limit $N\to\infty$.
The main object of interest is the single-particle Bloch Hamiltonian 
\begin{equation}\label{Eq: Bloch ham}
    H(k)=d_x(k)\sigma_x+d_y(k)\sigma_y,
\end{equation}
where $d_x(k)=v+w\cos ka$, $d_y(k)=-w\sin ka$ and $\sigma_i$ are Pauli matrices acting in the $A,B$ space.
The two energy bands are given by
\begin{align*}
    E_{\pm}&=\pm\sqrt{d_x^2+d_y^2}, \\
           &=\pm\sqrt{v^2+w^2+2vw\cos ka}~.
\end{align*}
The two bands can touch and close the gap when $v=\pm w$, at $ka=0$ $(v=-w)$ and $ka=\pi$ $(v=w)$. This signals a change between topologically distinct phases \cite{Asboth2016AInsulators}.  
\subsection{Symmetries and Topology}

The SSH model possesses two symmetries: sublattice symmetry $A\leftrightarrow B$, characterized by
\begin{equation*}
    \sigma_zH(k)\sigma_z=-H(k),
\end{equation*}
and inversion symmetry $j\to N-j+1$, characterized by 
\begin{equation*}
    \sigma_xH(k)\sigma_x=H(-k)~.
\end{equation*}
The former allows for the definition of a winding number invariant \cite{Chiu2016ClassificationSymmetries}, while the latter forces the quantization of this winding number to just two values \cite{Zak1989BerrysSolids}.
For a sublattice symmetric gapped Hamiltonian, the following equation defines a winding number \cite{Chiu2016ClassificationSymmetries}
\begin{equation}\label{Eq: Winding number Sublattice Symmetry}
\begin{split}
        \nu&=\int_{-\pi/a}^{\pi/a}\frac{\dd k}{4\pi  i }\tr \left[\sigma_zH^{-1}(k)\partial_k H(k)\right].
\end{split}
\end{equation}
This formulation allows one to write \cref{Eq: Winding number Sublattice Symmetry} in terms of Green's functions \cite{Gurarie2011Single-particleInsulators}.
This will be used later to investigate the invariant's fate in the interacting system.
For the SSH chain, we show in \cref{App: Quantization of Polarization} that this invariant yields 
\begin{equation}\label{Eq: winding SSH}
    \nu =\left\{\begin{alignedat}{5}
    & 0, \ \ \ \text{if } \left|v\right|>|w|~, \\
               & 1, \ \ \ \text{if } \left|v\right|<|w|~.
\end{alignedat}\right.
\end{equation}\\
We can relate the winding number $\nu$ to the Berry phase $\phi$ of the system at half-filling,
\begin{equation*}
    \phi=\int_{-\pi/a}^{\pi/a} \dd k\bra{u_-(k)} i \partial_k\ket{u_-(k)},
\end{equation*}
where $\ket{u_-(k)}$ is the lower band's Bloch state. It is quantized by virtue of inversion symmetry to $\phi=0$ for $|v|>|w|$ and $\phi=\pi$ for $|v|<|w|$.  

According to the modern theory of polarization \cite{King-Smith1993TheorySolids}, the Berry phase is closely related to the position of Wannier centers inside unit cells and, as such, indicates the presence/absence of a ``topological'' polarization,
\begin{equation}\label{Eq: topological polarization}
    P=\frac{e\phi}{2\pi}\mod e=\left\{\begin{alignedat}{5}
    & 0, \ \ \ \text{if } \nu=0, \\
               & \tfrac{1}{2}, \ \ \ \text{if } \nu=1.
\end{alignedat}\right.
\end{equation}
The $\mod e$ always appears in the definition of absolute polarization. It accounts for the $2\pi$ ambiguity of the Berry phase and physically for the choice of a unit cell \cite{King-Smith1993TheorySolids}. This polarization manifests itself as a bulk-boundary correspondence: if the system has open boundaries, there will be half-quantized charges sitting at the boundary of the system. This, in turn, means that there will be an anomalous response to an external electromagnetic field \cite{Aihara2020AnomalousPhase,Moustaj2024AnomalousInsulators}.

\subsection{Interactions and the continuum limit}
As we will be interested in the low-energy physics of the topological phases, we shall first derive the continuum limit of the noninteracting model at low energies. Since these transitions happen at $ka=0$ and $ka=\pi$, for the sake of simplicity, we take all parameters to be positive and expand the Bloch Hamiltonian around $ka=\pi$ to linear order in $ka$, 
\begin{equation}\label{Eq: continuum SSH Hamiltonian k-space}
\begin{split}
        H(\pi+ka)&=\begin{pmatrix}0 & v - w -  i  wka \\ v - w  +  i  wka & 0\end{pmatrix}\\
    &=m_0\sigma_x+wka\sigma_y,
\end{split}
\end{equation}
where $m_0=v-w$ will serve as a bare mass parameter in the field theory description. Note that for the transition at $k=0$, we need to replace $w\to-w$. The real space Hamiltonian is obtained by replacing $k\to- i \partial_x$ (we set $\hbar=1)$, 
\begin{equation}\label{Eq: continuum SSH Hamiltonian}
    \mathcal{H} = \int \dd x\Psi^\dagger(x)\left[m_0\sigma_x  - i  v_\mathrm{F}\sigma_y\partial_x\right]\Psi(x),
\end{equation}
where we introduced the Dirac spinor 
\begin{align}\label{Eq: Dirac Spinor}
    \hat{\Psi}(x_j)= \frac{1}{\sqrt{a}}\begin{bmatrix}
       c_{A,j}\\
       c_{B,j}
       \end{bmatrix},
\end{align}
and dropped the $j$ index when taking the limit $a\to0$.
We have also introduced the Fermi velocity $v_F=wa$.
This leads to the interpretation of the sublattice structure as a pseudospin degree of freedom.
Using the gamma matrix representation $\gamma_0=\sigma_x$, $\gamma_1= i \sigma_z$, and  $\bar{\Psi}\equiv\Psi^\dagger\gamma_0$ together with the slash notation $\slashed{\partial} = \gamma_0\partial_t - \gamma_1\partial_x$, we have the usual free Hamiltonian and Lagrangian densities of a Dirac fermion,
\begin{align*}
    \mathcal{H}_0(x) & = \bar{\Psi}(x)\left( i  v_\mathrm{F}\gamma_1\partial_x+m_0\right)\Psi(x), \\
    \mathcal{L}_0(x) & = \bar{\Psi}(x)\left( i  v_\mathrm{F}\slashed{\partial}-m_0\right)\Psi(x).
\end{align*}
Because of the constrained spatial degrees of freedom, 1D systems are generically strongly interacting -- fermions cannot move past each other without interacting \cite{Giamarchi2003QuantumDimension}.
To incorporate this effect, we consider the simplest interaction term that respects the symmetries of the system: a nearest-neighbor repulsion between spinless fermions placed in the same cell (intercell coupling would also break sublattice symmetry),
\begin{align}\label{Eq: nearestneighb repulsion}
   \mathcal{H}_\mathrm{I}= 
     V_0\sum_{j=1}^{N} n_{A,j}n_{B,j},
\end{align}
where $n_{\alpha,j}=c^\dagger_{\alpha,j}c_{\alpha,j}$, with $\alpha=A,B$. Usually, the densities are shifted by half to ensure half-filling and to respect particle-hole symmetry. This can also be accommodated by adding a chemical potential term in the free Hamiltonian. However, these terms do not matter in the continuum theory that we will be considering and will usually be treated using typical prescriptions such as normal ordering.
Using the Dirac spinor \cref{Eq: Dirac Spinor}, the continuum version of the interaction can be written as  
\begin{align*}
     \mathcal{H}_I= \frac{g}{2}\int \dd x\left[\bar{\Psi}(x)\gamma_\mu\Psi(x)\right]^2,
\end{align*}
where 
\begin{equation*}
    g=\lim_{a\to0}\frac{V_0a}{2}.
\end{equation*}
Combining all terms, the Lagrangian density for the continuum limit of the interacting SSH model is given by 
\begin{equation}\label{Eq: Thirring Lagrangian}
    \mathcal{L} = \bar{\Psi}(x)\left( i  v_F\slashed{\partial}-m_0\right)\Psi(x)-\frac{g}{2}\left[\bar{\Psi}(x)\gamma_\mu\Psi(x)\right]^2,
\end{equation}
which is the Thirring model \cite{Thirring1958ATheory,Korepin1979DirectModel}.

\subsection{Generalization: extended SSH model}
\begin{figure}[!hbt]
    \centering
    \includegraphics[width=\columnwidth]{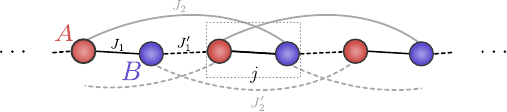}
    \caption{Extended SSH chain. Only the simplest extension has been considered in this figure, namely, the inclusion of $J_2$ and $J_2'$.}
    \label{fig: Extended SSH chain example}
\end{figure}
The SSH model can be generalized simply by including longer-range hoppings that respect all previously discussed symmetries. This model was studied in previous works to understand the interplay between long-range hoppings and disorder \cite{Perez-Gonzalez2019InterplaySystems} and, more recently, the effect of interactions on the topological phases \cite{Chang2024TopologicalModel}. An example of this extension to a next-next-nearest neighbor is shown in \cref{fig: Extended SSH chain example}. More generally, one can keep adding these hoppings by skipping $2n$ sites each time, resulting in the following Bloch Hamiltonian
\begin{equation}\label{Eq: Extended SSH model}
    H(k) = \sum_{n=1}^l\begin{pmatrix}
        0 &  h_n(k) \\ 
        h_n^*(k) & 0   
    \end{pmatrix},
\end{equation}
with $h_n(k) = J_ne^{- i  (n-1)k}+J_n'e^{ i  nk}$.
When $l=1$, this reduces to the regular SSH model with $J_1=v$ and $J_1'=w$. Because of its extended nature, higher winding numbers are possible, resulting in multiple edge state pairs. However, because the system still obeys inversion symmetry, the polarization \cref{Eq: topological polarization} is still quantized to $0$ (even winding numbers) or $1/2$ (odd winding numbers).

Consequently, this system has more possibilities for gap closings and transitions between two phases with different winding numbers. Nevertheless, as long as the interaction term \eqref{Eq: nearestneighb repulsion} is still the same, the Thirring model is still the low-energy theory that describes the physics near the transition point, but with modified Fermi velocities and bare masses. 
For example, we consider the model near the $k=\pi$ transition again. The bare mass and Fermi velocities are then given by 
\begin{equation}\label{Eq: Extended mass and velocity}
    \begin{split}
        m_0 &= \sum_{n=1}^l\left[(-1)^{n-1}J_n+(-1)^nJ_n'\right], \\
        v_\mathrm{F} &= a\sum_{n=1}^l\left[(-1)^{n-1}(n-1)J_n+(-1)^nJ_n'\right].
    \end{split}
\end{equation}

\section{Polarization as a response to an external field: non-interacting case} \label{Sec: Response Ext Field}

This section aims to show that we can formulate the electronic system's response to an external electric field by considering the effective action for a pure background gauge field using statistical field theory. 
The polarization density can be obtained by taking the derivative of the free energy with respect to an external electric field,
\begin{equation}\label{eq: pol density}
    \mathbf{P}=\frac{1}{V}\frac{\partial F}{\partial\mathbf{E}},
\end{equation}
where $V$ is the volume of the system. We will focus on the 1D case and derive formulas for the lattice and continuum versions of the SSH model.

\subsection{Lattice Model}
Let us start by considering the general expression of a non-interacting Hamiltonian on a lattice,
\begin{equation*}
    H=\sum_{ij}\sum_{\mu,\nu}c^\dagger_{i,\mu}h_{ij}^{\mu\nu}c_{j,\nu}=\sum_{k}\sum_{\mu\nu}c^\dagger_{k,\mu}h_{k}^{\mu\nu}c_{k,\nu}.
\end{equation*}
The indices $i,j$ refer to unit-cell, while $\mu,\nu$ refer to internal degrees of freedom, such as sublattice, orbitals, spins, etc. 
In a tight-binding formulation, the coupling to an external gauge field is achieved by performing a Peierls substitution, which stems from the principle of minimal coupling. That is, one replaces $h_{ij}^{\mu\nu}\to h_{ij}^{\mu\nu}e^{ i  A_{ij}}$, where $A_{ij}=\int_{x_i}^{x_j}A(x) \dd x$ is the integral of the gauge field from unit cell $i$ to $j$. Approximating the field to be constant on the scale of the unit cell and assuming weak coupling, we can linearize the exponential with respect to $A$. Subsequently, rewriting everything in momentum space yields
\begin{equation*}
    \begin{split}
        \begin{multlined}H=
            \sum_k\sum_{\mu\nu}c^\dagger_{k,\mu} h^{\mu\nu}_kc_{k,\nu} \\ \quad+\sum_{kq}\sum_{\mu\nu}c^\dagger_{k+q/2,\mu}\frac{\partial h^{\mu\nu}_k}{\partial k}c_{k-q/2,\nu}A_{-q},
        \end{multlined}
    \end{split}
\end{equation*}
where we used 
\begin{equation*}
    A_{ij} = \frac{1}{\sqrt{N}}\sum_qe^{ i  (x_i-x_j)q}A_q~.
\end{equation*}
We are also neglecting the diamagnetic contribution, proportional to $A^2$.
This approximation is justified when probing the polarization response of the system using a weak electric pulse.
With the full matrix Bloch Hamiltonian as $H(k)=\sum_{\mu,\nu}h_k^{\mu\nu}\ket{\mu}\bra{\nu}$, we can write down the imaginary time action as follows
\begin{equation*}
\begin{split}
        S[A]=\!\begin{multlined}[t]\int_0^{1/T} \dd \tau\sum_{kq}\Psi^*(k,\tau)\bigg\{\left[\partial_\tau-H\left(\frac{k+q}{2}\right)\right]\delta_{k,q} \\ -
    \partial_kH\left(\frac{k+q}{2}\right)A(q-k;\tau-\tau')\bigg\}\Psi(q,\tau),
    \end{multlined}
\end{split}
\end{equation*}
where the field $\Psi(k,\tau)$ is a Grassmann-valued field obtained by considering coherent states of the annihilation field operator in the second-quantization language. The integral over imaginary time $\tau$ runs from 0 to the inverse temperature $1/T$ (with units $\hbar=k_\mathrm{B}=1$). The partition function for this system is then conveniently written as a path integral over the fields $\Psi$ and $\Psi^*$, and has a formal solution
\begin{align*}
    Z[A]&=\int\mathcal{D}\Psi^*\mathcal{D}\Psi\exp\left(-S[A]\right) \\
    &=\exp\left[\Tr\ln\left(-G^{-1 }\right)\right],
\end{align*}
where we wrote $G^{-1}= G_0^{-1}+\mathcal{A}$, representing the Green's function as a formal operator, and $\Tr$ is a general trace operation over discrete and continuous indices. We have split the contributions into a ``bare'' one and one coming from the gauge field. The matrix $\mathcal{A}$ is the additional contribution from the derivative of the Hamiltonian. The ``bare'' Green's function (in Fourier space) can be obtained by finding the inverse of the following matrix
\begin{align*}
    G_0(k, i \omega_n)=\left[ i \omega_n+H(k)\right]^{-1},
\end{align*}
with fermionic Matsubara frequencies $\omega_n=\pi(2n+1)T$.
We now expand the matrix logarithm to linear order in $A$,
\begin{align*}
    Z[A]& \equiv  Z[0]\exp\left(-S_{\text{eff}}[A]\right),
\end{align*}
where the effective action is given by $S_{\text{eff}}[A]=\Tr\left(G_0\mathcal{A}\right)$. Working out the generalized trace yields
\begin{align*}
    \Tr\left(G_0\mathcal{A}\right) &= \sum_n\sum_q \Tilde{P}(q, i \omega_n)A(q, i \omega_n), \\
    \Tilde{P} (q, i \omega_n)&=\sum_k\text{tr}\left[G_0(k, i \omega_n)\partial_kH\left(k\right)\right]\delta_{q,0},
\end{align*}
where tr is the trace over internal degrees of freedom. Therefore, we can express the free energy as
\begin{equation*}
    F[A]=-T\ln(Z)=-T\ln(Z_0)+TS_{\text{eff}}[A].
\end{equation*}
Using \cref{eq: pol density} for the polarization density, we see that the functional derivative with respect to the electric field will only pick up the effective action contribution:
\begin{equation*}
    P(q, i \omega_n)= \frac{T}{V}\frac{\delta S_{\text{eff}}}{\delta E(q, i \omega_n)}.
\end{equation*}
We now choose a specific gauge where the applied electric field is constant and corresponds to a pulse.
For this, we use a gauge potential $A_x$ such that $E_x=-\partial_\tau A_x=E_0\delta(\tau)$. This is given by 
\begin{equation}
\label{Eq: Gauge Choice}
    A_x(x,\tau)=E_0T\sum_n\frac{e^{- i \omega_n\tau}}{ i \omega_n}.
\end{equation}
Hence, the electric polarization density becomes
\begin{equation}\label{Eq: lattice polarization qn}
    P(q, i \omega_n)=\frac{1}{\pi (2n+1)}\int_{-\pi}^\pi \frac{\dd k}{2\pi  i }\text{tr}\left[G_0(k, i \omega_n)\partial_kH(k)\right],
\end{equation}
where we have turned the sum into an integral over the Brillouin zone. The ``static'' version thereof, at zero temperature, is given by 
\begin{equation}\label{eq: PiPol}
    \mathcal{P}_{\mathrm{tot},0}\equiv \pi P(0,0)=\int_{-\pi}^\pi \frac{\dd k}{2\pi  i }\tr\left[G_0(k,0)\partial_kG_0^{-1}(k,0)\right],
\end{equation}
where we included the subscript $0$ in $\mathcal{P}_{\mathrm{tot},0}$ to indicate the non-interacting case. We could write \cref{eq: PiPol} because the Green's function at zero temperature is equivalent to the inverse of the Bloch Hamiltonian. 
Note that this type of polarization is not well defined for crystalline insulators, which is why the modern theory of polarization was developed in the first place \cite{King-Smith1993TheorySolids}. This occurs because there is a gauge degree of freedom in the choice of unit cells to define the dipole moment. 
However, the formulation of Green's functions is still beneficial. For a sublattice symmetric Hamiltonian, the relative polarization between sublattices $\mathcal{P}_0$ is a well-defined and gauge-invariant quantity, which we can obtain simply by multiplying the expression inside the trace by $\sigma_z/2$. \cref{eq: PiPol} then becomes
\begin{equation}\label{Eq: General Exp for Pol Winding}
    \mathcal{P}_0=\int_{-\pi}^\pi \frac{\dd k}{4\pi  i }\text{tr}\left[\sigma_zG_0(k,0)\partial_kG_0^{-1}(k,0)\right],
\end{equation}
which is equivalent to the winding number, \cref{Eq: Winding number Sublattice Symmetry}. Similar expressions have been discussed in Refs.~\cite{Gurarie2011Single-particleInsulators,Sbierski2018TopologicalApproach,Balabanov2022QuantizationTemperatures}, but we derive ours in a different context here. We further note that there is a factor two difference with the Berry phase definition of polarization and that this formulation does not take into account the$\mod{e}$ ambiguity. Since we are mainly interested in the continuum formulation, this should not be a problem as it still indicates the two topologically distinct phases, as we will see shortly.  

\subsection{Continuum polarization of the SSH model}
From this point onward, we shall set $v_F=1$.
From the continuum Hamiltonian of the SSH model \cref{Eq: continuum SSH Hamiltonian}, we can infer what the action functional of this system will be in the imaginary time formulation:
\begin{equation*}
\begin{split}
    S&=\int \dd x \dd \tau\Psi^*(x,\tau)\left[\partial_\tau- i \sigma_y\partial_x-m_0\sigma_x\right]\Psi(x,\tau) \\
    &=-\int \dd x \dd \tau\int \dd x' \dd \tau'\Psi^*(x,\tau)G_0^{-1}(x,\tau;x',\tau')\Psi(x',\tau'),
\end{split}
\end{equation*}
where we introduced the Green's function $G_0$, whose momentum representation is 
\begin{equation*}
\label{Eq: Free GF}
\begin{split}
     G_0(k, i \omega_n)&=\frac{- i \omega_n+k\sigma_y+m_0\sigma_x}{\omega_n^2+m_0^2+k^2}.
\end{split}
\end{equation*}
\newline
Instead of the Peierls substitution used in the lattice formulation, we will directly introduce a background gauge field through minimal coupling. That is, we replace $\partial_j\to\partial_j+ i  A_j$, where $A_\tau (x,\tau)=- i \phi(x,\tau)$, with $\phi(x,\tau)$ the electric potential, and $A_x(x,\tau)$ is the magnetic potential. This modifies the Green's function to $G^{-1}=G_0^{-1}-\mathcal{A}$, where
\begin{equation*}
    \mathcal{A}= -A_\tau+\sigma_yA_x.
\end{equation*}
Note that we are omitting coordinates for simplicity. We now integrate out the fermionic degrees of freedom in the partition function to obtain an effective action, which, to linear order, is given by  $S_{\text{eff}}[\mathcal{A}]=\Tr\left(G_0\mathcal{A}\right)$. 
The generalized trace operation is given by 
\begin{widetext}
\begin{equation}\label{Eq: Gen Trace Cont}
\begin{split}
    \Tr\left(G_0\mathcal{A}\right)&=\int \dd x \dd \tau\int \dd x' \dd \tau'\tr\left[G_0(x-x',\tau-\tau')\mathcal{A}(x',\tau'-\tau)\delta(x-x')\right] \\
    &=\int \dd \tau \dd \tau'\tr\left[G_0(0,\tau-\tau')\int \dd x\mathcal{A}(x,\tau'-\tau)\right].
\end{split}
\end{equation}
\end{widetext}
We choose a gauge potential in \cref{Eq: Gauge Choice} that yields an electric pulse at time $\tau$. We can then write \cref{Eq: Gen Trace Cont} as
\begin{equation*}
    \Tr\left(G_0\mathcal{A}\right)=T\sum_n\int\frac{\dd k}{2\pi}\tr\left[G_0(k, i \omega_n) i \sigma_y\right]A(0, i \omega_n),
\end{equation*}
Noting that $\partial_kG_0^{-1}(k, i \omega_n)=-\sigma_y$,
we have
\begin{equation*}
        \Tr\left(G_0\mathcal{A}\right)=T\sum_n\int\frac{\dd k}{2\pi  i }\tr\left[G_0\partial_kG_0^{-1}\right]A(0, i \omega_n).
\end{equation*}
This means we have the same expression for the polarization as in the lattice model, \cref{Eq: lattice polarization qn}. A direct calculation of the relative polarization, with $\sigma_z/2$ inside the trace, yields (see \cref{App: Quantization of Polarization}),
\begin{equation}\label{Eq: Continuum nonint polarization dens}
    \mathcal{P}_0=\frac{1}{2}\text{sign}\left(m_0\right).
\end{equation}
The half-quantization condition above arises due to the integration over momentum space covering the entire real line. Consequently, it is no longer a winding number, and there is no reason for it to be an integer. Nonetheless, it remains valuable in identifying the topological phase transition. Additionally, the continuum polarization can effectively capture the integer change in polarization across the transition, specifically 
$\Delta\mathcal{P}_0=1$ (in units of the free electron charge).

\section{Effects of Interactions}\label{Sec: Polarinteraction}

We have so far seen that the polarization density for a non-interacting system can be expressed, in the lattice and continuum theories, in terms of the noninteracting Green's function. We have used an imaginary time formulation to derive those expressions. Still, since we are interested in the static polarization, which does not contain any Matsubara frequency, we might as well consider a generic field theory instead of an Euclidean one. Equipped with this knowledge, we will now explore the effects of interactions in a more general setting. 

\subsection{Expectations from scaling properties}\label{Sec: Expectations from scaling}

Before we compute the polarization explicitly, here we present arguments as to its expected behavior in the case of the Thirring model \cref{Eq: Thirring Lagrangian}, which is an integrable massive quantum field theory (discussed in more detail in \cref{App: Thirring and Sine-Gordon models}).

Consider the full Green's function (including non-symmetry-breaking interactions that only contribute to the off-diagonal matrix elements) in its most general form,
\begin{equation}\label{eq: Gen GF f(k)}
    G(k,0)= \begin{pmatrix}0 & f(k) \\ f^*(k) & 0\end{pmatrix},
\end{equation}
where
\begin{equation*}
    f(k)=|f(k)|e^{ i \varphi(k)}.
\end{equation*}
By substituting \cref{eq: Gen GF f(k)} into \cref{Eq: General Exp for Pol Winding}, we obtain the polarization
\begin{equation}
\label{Eq: Polarization Int}
    \mathcal{P} = \lim_{\Lambda\to\infty}\frac{1}{2\pi}\int_{-\Lambda}^{\Lambda}\dd k\partial_k\varphi(k),
\end{equation}
where we have introduced a cutoff $\Lambda$, which we will come back to later. 
In other words, the polarization is a winding of the Green's function's complex phase $\varphi$ from $k=-\infty$ to $k=\infty$. 
As such, its contributions can only come from zeros or poles. 
Since the massive Thirring model, given by \cref{Eq: Thirring Lagrangian}, is integrable, we can safely assume that it is continuous and differentiable everywhere on the real axis, except possibly at $|k|=0,\infty$ (see Appendix \ref{App: Thirring and Sine-Gordon models}). 
Moreover, we also know from the Lehmann representation that the poles of the Green's function still correspond to excitation energies, even in the interacting case where they are renormalized.
Assuming that these are nonzero (since the theory is massive), the only contribution to the winding number would come from possible zeros at the IR and UV points $|k|=0,\infty$. 
It is thus sufficient to study the behavior of the Green's function at the fixed points of the renormalization group (RG) flow (IR and UV regions). The UV behavior of any theory can be accessed from the conformal field theory (CFT) at the UV point. 
If this theory is connected to a massive IR fixed point through the RG flow, then the Green's function behaves as follows:
\begin{equation}
        \label{Eq: Green's Function scaling massive}
        g(r)\sim\left\{\begin{alignedat}{5}
            & r^{-y}, \ \ \ \text{if } r\to 0~,  \\
            &  e^{-mr}, \ \ \ \text{if } r\to\infty~,
        \end{alignedat}\right.
    \end{equation}
where $y$ is the CFT scaling dimension of the field and $m$ is the mass of the lightest particle. In momentum space (see \cref{App: Derivation} for a derivation), this becomes
\begin{equation}\label{Eq: g(k) large small k}
    g(k)\sim\left\{\begin{alignedat}{5}
        & k^{y-2}, \ \ \ \text{if } k\to\infty~,  \\
        &  \frac{1}{m}, \ \ \ \text{if } k\to 0~.
    \end{alignedat}\right.
\end{equation}
Because of the mass gap, the topological invariant is well defined, as the integral \cref{Eq: Polarization Int} does not contain any divergence and can be written as
\begin{equation*}
    \mathcal{P} = \frac{1}{2\pi}\lim_{\Lambda\to\infty}\left[\varphi(\Lambda) - \varphi(-\Lambda)\right].
\end{equation*}
The only relevant contribution will be from negative large momenta, since then an extra phase $e^{\pm i\pi y}$ arises, as can be inferred from \cref{Eq: g(k) large small k}.
Hence, the polarization must be of the form $\mathcal{P} =\pm y/2$, for $y<1$. 

\begin{figure}[!t]
    \centering
    \includegraphics[]{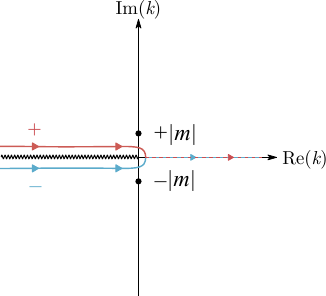}
    \caption{Different integration paths: above (red) or below (blue) the branch cut, depending on the sign of the mass.}
    \label{fig: branch cut path}
\end{figure}

The main conclusion is that $y$, the \textit{scaling dimension} of the field, redefines the amplitude of the continuum polarization. As for the sign, which determines the topological phase transition, it is still fully determined by the mass parameter,
\begin{equation}\label{Eq: polarization with scaling y}
    \mathcal{P} = \frac{1}{2}\text{sign}\left(m\right)y = y\mathcal{P}_0~,
\end{equation}
where $\mathcal{P}_0$ is the non-interacting polarization density from \cref{Eq: Continuum nonint polarization dens}. The reason for this is the following [see \cref{fig: branch cut path}]: the integration runs along the real axis, and the winding function has a branch cut parallel to it, with branching points at infinity and at the Green's function's zero. The choice of Riemann sheet for either $e^{i\pi}$ or $e^{-i\pi}$ is crucial and is enforced by the mass parameter. It dictates the direction from which the branch cut is approached, i.e., the Riemann sheet of the winding function over which the integration runs.
If the mass gap vanishes, the integration runs precisely over the branch cut and does not produce a well-defined result. This is consistent with the fact that the polarization is not defined for a gapless system. 

These assumptions can be explicitly verified in the free case, when $g(k)=e^{ i \varphi(k)}$, with 
\begin{equation*}
    \varphi(k) =\arctan\left(\frac{-k}{m}\right).
\end{equation*}
Since $\varphi(k)$ has a logarithmic branch cut, the Riemann surface contains an infinite number of different sheets. However, physically, only the difference between two adjacent sheets is relevant. This result holds for any chiral invariant model, emphasizing the importance of a mass gap to protect the topological phase.

\subsection{Polarization for the Thirring model}
Our next goal is to use insights about the Thirring model to derive exact results for the polarization \cref{Eq: General Exp for Pol Winding}, using the interacting Green's function. As shown in the previous section, one may expect to find all the information in the scaling dimension of the fermionic field, provided that the system is gapped. The Thirring Lagrangian, given by \cref{Eq: Thirring Lagrangian}, has two parameters: a bare mass $m_0$ (which is just the dimerization of the chain) and the local self-interaction coupling strength $g$, considered strictly positive in our analysis.
However, since this is a quantum field theory, the physical mass $m$ and \textit{effective coupling} $B$ are not equal to the respective bare quantities from the Lagrangian.
To extract non-perturbatively the relations between physical and bare parameters of the field theory, we can exploit the mapping between the Thirring and the sine-Gordon model \cite{Coleman1975QuantumModel} (for more details, see \cref{App: Thirring and Sine-Gordon models}). The Lagrangian of the latter is
\begin{equation*}
    \mathcal{L}_\mathrm{SG} = \frac{v^2_\mathrm{F}}{16\pi}\left(\partial_\mu\Phi\right)^2 + \frac{m_0^2}{\beta^2}\left(\cos{\beta\Phi}-1\right)~,
\end{equation*}
where $\Phi(x,t)$ is a bosonic scalar field, and the effective coupling is
\begin{equation*}
    \beta^2 = \frac{1}{2}\frac{1}{1-g/\pi}~.
\end{equation*}
The sine-Gordon model describes a theory of stable solitonic excitations, which interpolate between different classical vacua of the theory $\Phi_{0,n}=2n\pi/\beta$. This correspondence allows one to link the fermionic fields $\Psi$ of the Thirring model to the operators in the sine-Gordon model that create or annihilate solitons, $e^{ i  \beta\Phi/2}\mathcal{O}^1_{\beta/2}$. Here, $\mathcal{O}^1_{\beta/2}$ is a string operator, which accounts for the symmetry among the classical vacua $\Phi\to\Phi+2n\pi/B$. This symmetry is also present in the quantum theory.

As an example of this mapping, it was shown \cite{Zamolodchikov1995MassReductions} that the physical mass of Thirring fermions and sine-Gordon solitons scales as a power law of the bare mass
\begin{equation}
    \label{Eq: Physical Mass}
    m^2(m_0,g) = C(g)\left(m_0^2\right)^{\frac{1}{2-2\beta^2}},
\end{equation}
where $C(g)$ is a function of the bare Thirring coupling $g$.
This exact result was obtained by using non-perturbative bosonization methods in Ref.~\cite{Zamolodchikov1995MassReductions} (see \cref{App: Thirring and Sine-Gordon models} for explanations).
Additionally, the repulsive regime of the Thirring model, $\pi/2>g>0$, corresponds to the regime $1/2<\beta^2<1$ in the sine-Gordon model. In this case, Thirring fermions and antifermions, carrying a renormalized electronic charge $e^*$, are mapped into solitons and antisolitons, carrying a topological charge. The latter is defined as
\begin{equation}
    \label{Eq: Topo Charge}
    Q_t = \frac{\beta}{2\pi}\int_{-\infty}^{+\infty}\dd x\partial_x\Phi(x,t).
\end{equation}
Note that, in this case, the term topological refers to how the field $\Phi(x,t)$ interpolates between the different minima of the sine-Gordon potential, i.e., the symmetry between different ground states of the model, and not to the topological phase. 

Exploiting this mapping even further, the zero frequency Green's function of the Thirring model, defined as
\begin{equation}
    \label{Eq: Green's function Thirring}
    G(k) = \int_{-\infty}^{+\infty}\frac{\dd x}{2\pi}e^{ i  kx} \langle\Psi(x)\bar{\Psi}(0)\rangle ,
\end{equation}
can be computed in terms of the form factors of the soliton-creating operator of the sine-Gordon model, $\tilde{\mathcal{O}} = e^{ i  \beta\Phi/2}\mathcal{O}^1_{\beta/2}$,
\begin{equation*}
    F_n^{\tilde{\mathcal{O}}}(\theta_1,\cdots,\theta_n)\equiv\bra{0}\tilde{\mathcal{O}}\ket{\theta_1,\cdots,\theta_n},
\end{equation*}
where $\theta_i$ are the rapidities of particles related to their momenta via $p_i = m\sinh\theta_i$ (see \cref{App: Thirring and Sine-Gordon models} for more details and the soliton-creating operator's exact form). The Green's function then takes the form
\begin{equation}
    \label{Eq: SGFFGreensfunction}
    \begin{split}
            G(k) = \sum_{n=0}^\infty\int_{-\infty}^{+\infty} & \frac{\dd x}{2\pi} \frac{\dd\theta_1}{2\pi}\dots\frac{\dd\theta_n}{2\pi}e^{ i  kx}|F_n^{\tilde{\mathcal{O}}}(\theta_1,\dots,\theta_n)|^2\\
    &\times\delta\left(k-m\sum_{j=1}^n\sinh\theta_j\right).
    \end{split}
\end{equation}
Using the UV-normalization proposed in Ref.~\cite{Lukyanov2001FormModel},
the scaling dimension of the Thirring fermionic field (which is the exponent of the power-like scaling at large momenta, discussed in \cref{Sec: Expectations from scaling}) is found to be
\begin{equation}
    \label{Eq: Scaling dim fermion}
    y = \frac{1}{2}\left(\frac{1}{1-g/\pi} + 1-\frac{g}{\pi}\right).
\end{equation}
This result can be numerically verified from \cref{Eq: SGFFGreensfunction} using the Feynman gas method \cite{Mussardo2020FormTheory}.
The scaling of \cref{Eq: Green's function Thirring} in the UV region is also consistent with conformal perturbation theory \cite{Zamolodchikov1989IntegrableTheory}.
The exponent $y$ can be identified as the scaling dimension of the field in the related CFT (more details are available in \cref{App: Scaling dimension}).

Therefore, the polarization density of the interacting SSH model in the continuum limit may be readily computed using \cref{Eq: polarization with scaling y} and \cref{Eq: Scaling dim fermion},
\begin{equation}
    \label{Eq:Polarization res TM}
    \mathcal{P} = \frac{1}{4}\left(\frac{1}{1-g/\pi} + 1-\frac{g}{\pi}\right)\text{sign}(m)~.
\end{equation}
This is the main result of this paper. However, there are limitations.

It is known that when the interaction is strong enough (i.e., $g>\pi/2$), the system becomes gapless \cite{Coleman1975QuantumModel}.
In this case, the interaction makes the massive term irrelevant in the RG sense.
It does not drive the system out of the critical region but explores it, as the Hamiltonian is neither gapped nor conformal.
The physics of this scaling region is closer to the Luttinger liquid and is not captured by the behavior of the polarization, which is well defined for gapped systems as insulators.
Additionally, an important property of the physical mass is that the gap is closed when the effective coupling $B$ reaches an essential singularity (i.e., $\beta^2=1$ or $g = \pi/2$) of \cref{Eq: Physical Mass}.
Otherwise, the sign of the mass never changes.
This means that the polarization jumps along the same line $m_0 = 0$ of the free case, and the effect of interactions does not change the shape of the phase diagram in this massive regime. 
\begin{figure}[!t]
    \centering
    \includegraphics[width=\columnwidth]{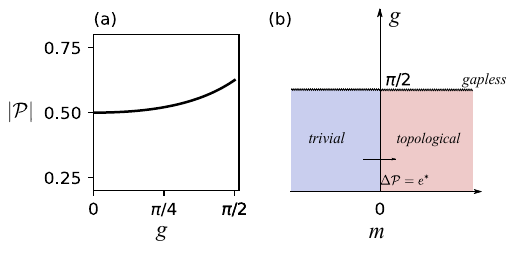}
    \caption{(a) Continuum polarization as a function of the interaction strength of the Thirring model. The deviation from the noninteracting value is relatively small in the region where the interaction strength allows for a well-defined polarization, i.e., $0\leq g<\pi/2$. (b) Phase diagram for the low-energy sector of the repulsive interacting SSH chain in the region where the polarization is well defined. We identify trivial and topological regions from the exact mapping established at the level of the free theory (i.e., along the $m$-axis). The introduction of interaction in the system does not spoil the topology of the chain until the interaction overcomes the threshold (represented by the wiggling line) and the system undergoes a phase transition. The jump in the polarization, here represented by a change of color from blue to red, along the $g$-axis, signals the passage from a trivial into a topological phase, accompanied by a change in polarization $\Delta\mathcal{P}=e^*$, where $e^*$ is the charge carried by the excitations of the theory.}
    \label{fig: phase diagram}
\end{figure}

The polarization of the continuum limit of the interacting SSH model and the corresponding phase diagram are shown in Figs.~\ref{fig: phase diagram}(a) and \ref{fig: phase diagram}(b), respectively. We note that the phase diagram in the continuum limit is different from that of the lattice model.
This is because we are considering different approaches: we are studying the polarization in the continuum limit using field theoretical tools, while most numerical studies consider the behavior of different order parameters on the lattice.
This means the continuum limit does not capture boundaries, finite-size effects, and excited state properties of lattice models, such as band curvature. It might fail to determine what other order parameters can be captured , such as the different shapes of critical lines in the original lattice model.
For instance, the \textit{gapless} straight line in \cref{fig: phase diagram}(b) is different from the one obtained in the literature, where numerical schemes were used to compute similar phase diagrams \cite{Nersesyan2020PhasePoint,Zhou2023ExploringModel}. 

One of the main interpretations of these results is that the polarization is only modified to account for the fractionalized charge of the excitations of the theory.
Indeed, the term $(1-g/\pi)^{-1}$ and its inverse, appearing in \cref{Eq:Polarization res TM}, come from two contributions: the renormalized electronic charge of the Thirring fermion and a second term that is a consequence of the degeneracy of the ground state, carried by the string operator accompanying the soliton creation operators, respectively.
Thus, electron-electron interactions do not change the qualitative description of the topological phases but do quantitatively modify it to account for the renormalization of the polarization. 
The situation is somewhat similar to the study of the topological invariant in the presence of interactions in the integer quantum Hall effect. In this case, it has been shown that it does not drive a topological phase transition \cite{Zhang2022InfluenceEffect}.

Moreover, if one were to construct an adiabatic charge pump, such as the Rice-Mele charge pump \cite{Asboth2016AInsulators}, the quantized charge transferred after one cycle would now be fractionalized to this new value.
Indeed, the induced current results from a change in polarization, driven by a change in configuration in the insulator's bound charges.
The Rice-Mele pump is known to adiabatically modulate a 1D chain, such that it is tuned from the trivial to the topological phase of the SSH model and back again to its trivial phase \cite{Asboth2016AdiabaticModel}.
As a result, the change in polarization is $\Delta P = 1$ (recall that we are working with units where $e=1$).
This picture is still valid in the continuum model, with $\Delta \mathcal{P}=1$. As such, we can interpret the modification of the amplitude of the polarization in \cref{Eq:Polarization res TM} as a result of the modification of the charge of the elementary excitations of the theory, i.e., $\Delta \mathcal{P} = e^*$, as sketched in \cref{fig: phase diagram}.

Finally, we would like to point out that all the results derived in the last section also hold when one considers the generalization to the extended SSH model \cref{Eq: Extended SSH model}. Even though the winding number of the lattice theory might take higher values, signaling multiple topologically distinct phases, the polarization itself is only nontrivial whenever the winding number is an odd integer [see \cref{Eq: topological polarization}, which is defined modulo the uncertainty quantum $e$]. The general case contains infinitely many topologically distinct atomic limits, half of which are of the obstructed nature \cite{Bradlyn2017TopologicalChemistry}. When taking the continuum limit near a phase transition, the Thirring model describes two topologically distinct phases corresponding to neighboring winding numbers. In other words, it models the transition between a phase with nontrivial polarization and one with zero polarization. The only change occurs in different definitions of the Fermi velocities $v_F$ and mass parameters $m_0$, which are given in \cref{Eq: Extended mass and velocity}.

To gain a deeper understanding of these results, we will employ perturbation theory, the renormalization group, and terminology from Fermi liquid theory to provide additional physical insights and interpretations.

\subsection{Perturbation theory}
Perturbation theory requires regularization schemes, such as introducing a cutoff $\Lambda$ in momentum space, which we can consider proportional to the inverse of the lattice spacing in the related fermionic chain problem.
Such a cutoff $\Lambda$ is needed when the interaction parameter $g$ is considered because perturbation theory produces divergent contributions. These are resolved either by considering counterterms in the Lagrangian or are regularized by the presence of the cutoff $\Lambda$ and eliminated by a convenient redefinition.
In practice, an infinitesimal redefinition of the cutoff $\Lambda\to\Lambda+\delta\Lambda$ can be performed, leading to a shift in all other parameters of the theory subject to a renormalization, i.e., the coupling constants and the field prefactors,
\begin{eqnarray*}
    &g_i\to g_i+\delta g_i~, \\
    &\Psi\to (1 +\delta\eta)\Psi~, \\
    &\beta^{(i)}_\mathrm{RG} \equiv -\Lambda\frac{\delta g_i}{\delta\Lambda}\simeq -\frac{\partial g_i}{\partial \ln\Lambda}~, \\
    &\gamma = -\Lambda\frac{\delta\eta}{\delta\Lambda}\simeq -\frac{\partial \eta}{\partial \ln\Lambda}~,
\end{eqnarray*}
where $i=1,2$, with $g_1=m$ and $g_2=g$. We have also defined the RG beta functions $\beta_\mathrm{RG}$ of the parameters $g_i$ and implied a change in the scaling dimension $\gamma$ because of the redefinition of the field through $\eta$. Additionally, the correlation functions containing $n$ fields transform as
\begin{equation*}
    G^{(n)}(\Lambda,g_i)\to (1+n\delta\eta)G^{(n)}(g_i)~.
\end{equation*}
Requiring that their form must be independent of the value of the cutoff $\Lambda$ leads to the so-called Callan-Symanzik equation:
\begin{equation}\label{Eq: CSequation}
\begin{split}
        \frac{\dd}{\dd\ln\Lambda}G^{(n)}(k_1,\dots,k_n;\Lambda,g) &= 0 \\
    \left[\Lambda\frac{\partial}{\partial\Lambda} +\sum_i\beta_\mathrm{RG}(g_i)\frac{\partial}{\partial g_i} + n\gamma\right]G^{(n)}(k_1,\dots,k_n;\Lambda,g)  &= 0 .
\end{split}
\end{equation}
Using the fact that $\beta_\mathrm{RG}\to0$ in the UV limit, \cref{Eq: CSequation} results in $\ln{G}\sim 2\gamma\ln(k/\Lambda)$. Therefore, we find that the argument of the two-point function is
\begin{equation*}
    \varphi(k)\sim- i \ln G = -2 i \gamma\ln\left(\frac{k}{\Lambda}\right),
\end{equation*}
which corroborates the dependence of the polarization on the scaling dimension of the fermionic field. Taking $k\to\Lambda$ and then $\Lambda\to\infty$ also confirms that the polarization is independent of regularization schemes and the form of the cutoff. 

Although perturbation theory in 1D systems differs significantly from higher-dimensional cases \cite{Giamarchi2003QuantumDimension}, it can provide valuable insights for interpreting results.
For instance, \cref{Eq: polarization with scaling y}, derived non-perturbatively, can be analyzed using Fermi liquid terminology. This approach paves the way for potential generalizations of our arguments to interacting topological fermionic systems in 2D or 3D. 
This can be shown by associating the renormalization group procedure with Fermi liquid theory. The simplest way to do so is by considering the many-body Green's function close to its poles, which takes the form
\begin{equation}
    G(k,\omega) = \frac{Z}{\omega - \epsilon(k)}~,
\end{equation}
where $\epsilon(k)$ is the energy of the excitations with momentum $k$. The shift $\delta\eta$, dictated by the scaling dimension of the fermion field, induces a shift of the quasiparticle weight $Z$.
In the Fermi liquid framework, this quantity measures Fermi-liquid behavior: as long as $ Z\simeq 1$, excitations behave as free electrons with dressed mass and electric charge, i.e., they are the system's quasiparticles.
When this is the case, we do not expect the polarization to be affected by interactions, except for the ``trivial'' renormalization of the electronic charge. However, our result \cref{Eq:Polarization res TM} contains an additional contribution related to the degeneracy of the ground state of the dual bosonic model, breaking the Fermi liquid picture and signaling the fact that the fermionic excitations are not reducible to dressed electrons but are entirely different.

\begin{figure}[!b]
    \centering
    \includegraphics{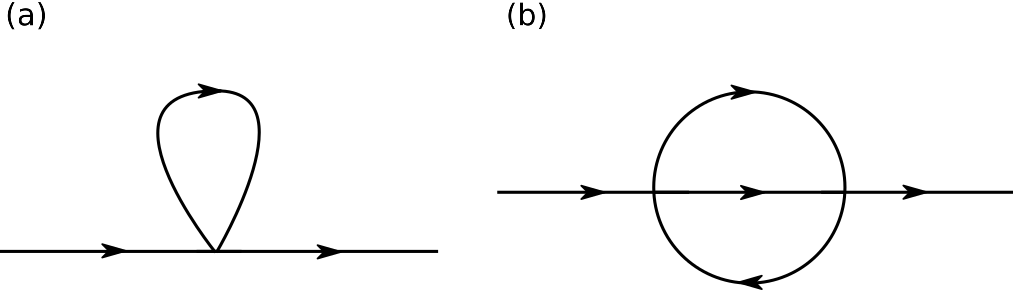}
    \caption{The two relevant corrections to the Green's function. (a) The self-energy is momentum-independent in this case. (b) The lowest correction to the self-energy acquires a linear momentum dependence, resulting in a $g^2$ correction to the polarization.}
    \label{fig: 1st selfenergy}
\end{figure}
Let us now compute the lowest order corrections to the polarization given by \cref{Eq: General Exp for Pol Winding}, but with the full interacting Green's function.
The polarization can be expanded as
\begin{equation*}
    \mathcal{P} = \mathcal{P}^{(0)} + g\mathcal{P}^{(1)} + \frac{g^2}{2}\mathcal{P}^{(2)} + \dots
\end{equation*}
The lowest order expansion of the exact solution of the polarization \cref{Eq:Polarization res TM} is of order $g^2$. Thus, we expect that perturbation theory reproduces this functional relationship, i.e., $\mathcal{P}^{(1)}=0$ and $\mathcal{P}^{(2)}\neq0$. 

To perform the calculations, we first recall that the fully interacting Green's function obeys the self-consistent Dyson equation, which is formally given by
\begin{equation}\label{Eq: Int GF Dyson}
    \begin{split}
        \hat{G}&=\hat{G}_0+\hat{G}_0*\hat{\Sigma}*\hat{G} \\
    &=\hat{G}_0+\hat{G}_0*\hat{\Sigma}*\hat{G_0}+\hat{G}_0*\hat{\Sigma}*\hat{G_0}*\hat{\Sigma}*\hat{G_0}+\cdots,
    \end{split}
\end{equation}
where the hat is used to denote the formal operators and the stars denote their multiplication. The self-energy $\hat{\Sigma}$ represents all the one-particle irreducible diagrams in the expansion of the two-point correlation function (i.e., the Green's function). The inverse of the Green's function is simply
\begin{equation}\label{Eq: Inv GF Int}
    \hat{G}^{-1}=\hat{G}_0^{-1}-\hat{\Sigma}~.
\end{equation}
Substituting \cref{Eq: Int GF Dyson} and \cref{Eq: Inv GF Int} into \cref{Eq: General Exp for Pol Winding}, one obtains two terms, which will be considered separately, 
\begin{equation}\label{Eq: Term 1}
    \Tr\left[\sigma_zG\partial_kG_0^{-1}\right] 
\end{equation}
and
\begin{equation}
    \label{Eq: Term 2}
    \Tr\left[\sigma_zG\partial_k\Sigma\right].
\end{equation}
To first order, the self-energy $\hat{\Sigma}$ does not depend on momentum (see the Feynman diagram in \cref{fig: 1st selfenergy}). 
\begin{figure}[!t]
    \centering
    \includegraphics[width=0.5\columnwidth]{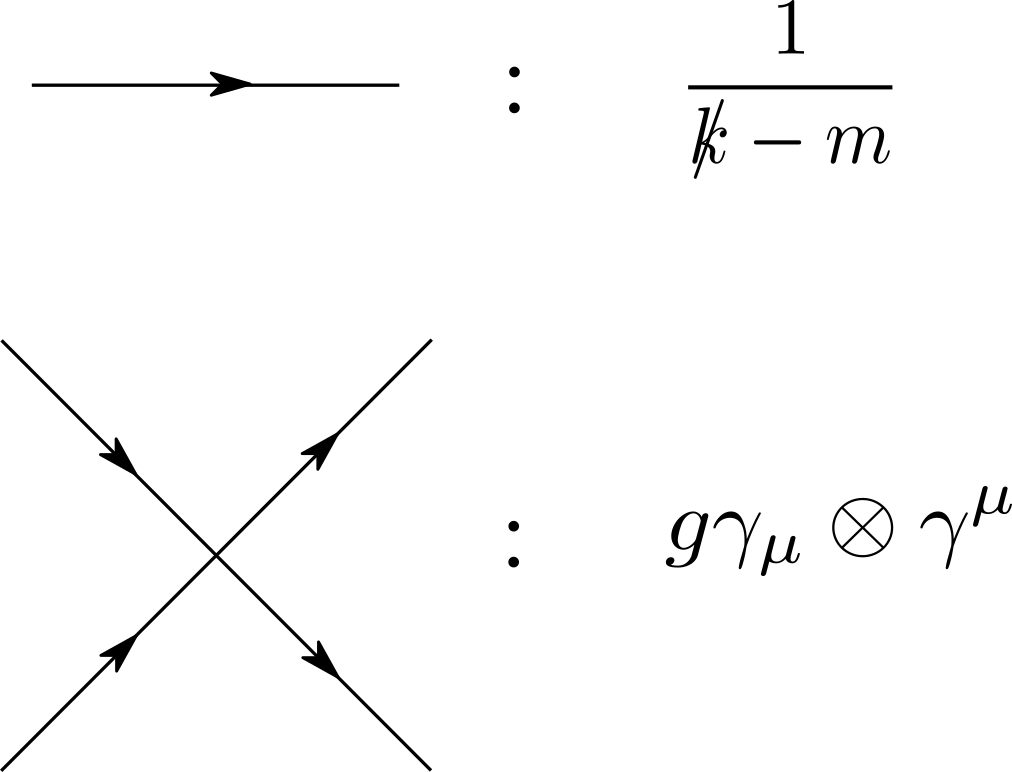}
    \caption{Feynman rules for the propagator in momentum space and the interaction vertex. The relation to the initially used propagator is obtained by multiplying it from the left by $-\sigma_x$.}
    \label{fig: Feynmanrules}
\end{figure}
Thus only the first term \cref{Eq: Term 1} contributes to the first-order correction to the self-energy, i.e.,
\begin{equation}\label{Eq: First order correction Polarization}
    g\mathcal{P}^{(1)} = \int_{-\infty}^{+\infty}\frac{\dd k}{4\pi  i } \Tr\left[\sigma_z G_0\Sigma G_0\partial_kG_0^{-1}\right].
\end{equation}

To compute $\Sigma$, we will use the fully relativistic propagators, vertex functions, and the Feynman rules depicted in \cref{fig: Feynmanrules}. 
The self-energy depicted in \cref{fig: 1st selfenergy}(a) then reads
\begin{align*}
    \Sigma^{(1)} &= g\int\frac{\dd^2q}{(2\pi)^2}\gamma_\mu G_0(k,\omega)\gamma^\mu \\
    & =2mg\mathbbm{1}\int \frac{\dd^2q}{(2\pi)^2}\frac{1}{\omega^2-k^2-m^2}~.
\end{align*}
Without proceeding with the calculation of this divergent integral (which requires a careful renormalization with a counterterm in the Lagrangian), we can nevertheless see, after multiplying by $-\sigma_x$ (see \cref{fig: 1st selfenergy}), that 
\begin{align}\label{Eq: selfenergy 1}
    \Sigma^{(1)} &= \alpha g\sigma_x~,
\end{align}
where $\alpha$ is the constant obtained by carrying out the integral. Substituting \cref{Eq: selfenergy 1} into \cref{Eq: First order correction Polarization}, we obtain
\begin{equation*}
    g\mathcal{P}^{(1)} = -g\alpha\int_{-\infty}^{+\infty}\frac{\dd k}{2\pi }\frac{(k^2-m^2)k}{(k^2+m^2)^2}~,
\end{equation*}
which is zero because the integrand is an odd function of $k$, confirming that there are no corrections to the free theory at first order in $g$.

The second-order contribution to the polarization comes from a term of the second type of first order contributions \cref{Eq: Term 2}. This means that we require a momentum-dependent self-energy correction, for which the simplest example is shown in \cref{fig: 1st selfenergy}(b) and given by 
\begin{equation*}
    \Sigma^{(2)}=g^2\int\frac{\dd^2q\dd^2q'}{(2\pi)^4}\gamma^\mu G_0(k+q-q')\gamma_\mu G_0(q)\gamma^\nu G_0(q')\gamma_\nu .
\end{equation*}
This expression is rather lengthy and leads to divergences if evaluated without proper regularization. For the sake of simplicity, we assume a linear dependence on momentum. In some limiting cases, this linear dependence can be derived explicitly (see Ref.~\cite{Lukyanov2001FormModel} for a specific example). Then, we take 
\begin{equation*}
    \Sigma^{(2)} =  \frac{g^2}{2}\sum_{\mu=0}^3\alpha_{\mu}\sigma_\mu k~,
\end{equation*}
where now $\sigma_0=\mathbbm{1}$, and $\sigma_i$, with $i=1,2,3$, are the usual Pauli matrices. The simplest term that can contribute to the polarization to second order is then
\begin{align*}
    \frac{g^2}{2}\mathcal{P}^{(2)} &= -\frac{g^2}{2}\int\frac{\dd k}{4\pi i}\Tr\left[\sigma_zG_0\partial_k\Sigma^{(2)}\right] \\
    &= \frac{g^2}{2}\int\frac{\dd k}{2\pi}\frac{\alpha_1k-\alpha_2m}{k^2+m^2}~.
\end{align*}
We can calculate the integral explicitly and obtain 
\begin{equation}
    \mathcal{P}^{(2)}= -\frac{\alpha_2}{2}~,
\end{equation}
confirming our expectations that the first correction in perturbation theory to the polarization should be of order $g^2$.

The absence of a first-order contribution also highlights that charge renormalization is not the only process contributing to the polarization response in the Thirring model  \cite{Korepin1993QuantumFunctions}.  If it were, a term proportional to $g/\pi$ would be present.
This supports the mathematical assumptions made using non-perturbative arguments and the renormalization group. From a physical perspective, it confirms that the breakdown of the Fermi liquid paradigm in 1D models also affects the topological properties related to their linear response.


\section{Conclusion} \label{Sec: Conclusion}
An analytic determination of topological properties of interacting fermionic systems has always proven to be a daunting task; so far, numerical results have been widely used in the field, but in this work, we have explicitly evaluated the topological invariant \cref{Eq: General Exp for Pol Winding} in the low-energy regime of the interacting SSH chain with nearest-neighbor density-density coupling.
The physical interpretation of this quantity was given in the free case as the relative polarization between sublattices.
The interaction term that we considered respects the symmetries of the non-interacting system, and its continuum limit yields the Thirring model, which enabled the formulation of the problem using this framework and allowed us to obtain exact analytic results.
We find that the IR and UV regions play a prominent role in the continuum theory.
Namely, the presence of a mass gap in the infrared regime protects the result obtained by the characterization of the ultraviolet regime in terms of the scaling dimension of the fermionic field.
The role of the mass parameter in determining the topological phase transitions of the lattice model is akin to how the interplay between the Semenoff mass and the Haldane mass dictates topological phase transitions in the Chern insulator \cite{Haldane1988ModelAnomaly}.
Our result can be readily generalized to cases where longer-range hoppings are included. Such models allow for more topological phase transitions. Still, in the vicinity of each of them, they can all be modeled by Thirring field theories where the details of different transitions are encoded by the various masses and the Fermi velocities \cref{Eq: Extended mass and velocity}.
The jump in the relative polarization is then computed in the same way as for the interacting SSH.

Moreover, its interpretation in terms of polarization and a quantized charge-pumping argument facilitates the description of topological phases through the bulk-boundary correspondence.
This enables the construction of a phase diagram within the applicable regime.
The modification of the polarization amplitude can be understood as coming from the renormalized charge, together with a quantity that weighs the degeneracy of the ground state carried by the excitations of the interacting theory. 
Although our results only apply to a family of 1D lattice models described by integrable interacting field theories in the continuum limit, they provide numerous insights regarding the nature of topological phase transitions in interacting models. For instance, the most relevant effects of interactions are the renormalization of the charge and, in this case, the degeneracy of the ground state. Additionally, most of the topological properties in the low-energy regime are determined by the scaling regions of the RG flow, where the UV regime influences the magnitude of the polarization \cref{Eq: polarization with scaling y} and the IR protects it via the presence of a gap. Finally, the non-Fermi liquid behavior leads to a simple redefinition of the polarization jump from integers to reals but does not affect the shape of the phase diagram.

As an outlook, it would be valuable to investigate whether this picture still holds when considering different types of interactions. The resulting low-energy effective field theories would vary, and some may not be integrable. Nevertheless, perturbative methods can always be used to compute the scaling dimensions of the fields and study the modification of the polarization.

\begin{acknowledgments}
    \noindent E.D., A.M., and C.X. are shared first authors.
    L.F., A.K.M., C.M.S., and D.S. supervised the project.

    \noindent A.M. acknowledges funding from the project TOPCORE with project No. OCENW.GROOT.2019.048, financed by the Dutch Research Council (NWO).

    \noindent C.X. acknowledges financial support from the National Research Fund Luxembourg under Grants No. C20/MS/14764976/TopRel.

    \noindent A.K.M. acknowledges
    financial support from Science Foundation Ireland
    through Grant 21/RP-2TF/10019.

    \noindent This work is part of the D-ITP consortium, a program of the Dutch Research Council (NWO) that is funded by the Dutch Ministry of Education, Culture and Science (OCW).
\end{acknowledgments}


 \bibliography{Refs} 

\appendix

\section{Quantization of the winding number} \label{App: Quantization of Polarization}

\subsection{Lattice model}
The Green's function for the Hamiltonian given by \cref{Eq: Bloch ham} is 
\begin{equation}\label{Eq: Green's function Lattice}
    G_0(k,0)= \frac{1}{(v^2+w^2+2vw\cos k)}\begin{pmatrix}0 & v+we^{- i  k} \\ v+we^{ i  k} & 0\end{pmatrix}.
\end{equation}
Working out the trace in \cref{Eq: General Exp for Pol Winding} yields
\begin{equation*}
    \nu=\int_{-\pi}^\pi\frac{\dd k}{4\pi  i }\frac{2 i  w \left(w + 
   v \cos k\right)}{v^2+w^2+2vw\cos k}.
\end{equation*} 
We will solve this using contour integration. Before we do that, we shall rewrite the equation as 
\begin{align*}
    \nu=\int_{-\pi}^\pi\frac{\dd k}{4\pi }\frac{2 + 
   V e^{ i  k}+Ve^{- i  k}}{(1+Ve^{ i  k})(1+Ve^{- i  k})},
\end{align*}
where we have introduced the parameter $V=v/w$. This integral can be rewritten as an integral over the unit circle $S^1$ in the complex plane by defining $z=e^{ i  k}$,
\begin{align*}
    \nu&=\oint_{S^1}\frac{\dd z}{2\pi i }\frac{1}{2V}\frac{V z^2+2z 
   +V}{z(z+V^{-1})(z+V)}, \\
    &\equiv\oint_{S^1}\frac{\dd z}{2\pi i }f(z).
\end{align*}
The function $f(z)$ has two poles inside the unit circle and one outside. The pole $z_0=0$ is always inside, while $z_1=-V^{-1}$ is inside if $|w|<|v|$ (making $z_2=-V$ outside). Using the residue theorem, we then have the following conditions
\begin{equation*}
\mathcal\nu=\left\{\begin{alignedat}{5}
    & \text{Res}(f,z_0)+\text{Res}(f,z_1), \ \ \ \text{if } |w|<|v| , \\
               & \text{Res}(f,z_0)+\text{Res}(f,z_2), \ \ \ \text{if } |w|>|v| .
\end{alignedat}\right.
\end{equation*}
Since the poles are of order 1, the residues are simply given by 
\begin{align*}
    \text{Res}(f,z_0)&=\lim_{z\to z_0}(z-z_0)f(z)=\frac{1}{2}, \\
    \text{Res}(f,z_1)&=\lim_{z\to z_1}(z-z_1)f(z)=-\frac{1}{2}, \\
    \text{Res}(f,z_2)&=\lim_{z\to z_2}(z-z_2)f(z)=\frac{1}{2},
\end{align*}
which finishes the proof of the quantization of the winding number,
\begin{equation*}
\nu=\left\{\begin{alignedat}{5}
    & 0, \ \ \ \text{if } |w|<|v|  ,\\
               & 1, \ \ \ \text{if } |w|>|v|.
\end{alignedat}\right. 
\end{equation*}

\subsection{Continuum model}\label{AppA: Continuum model}
In this case, we start from \cref{Eq: continuum SSH Hamiltonian k-space} and write down the Green's function as 
\begin{equation}\label{Eq: Green's function continuum}
    G_0(k,0)= \frac{1}{m^2+w^2k^2}\begin{pmatrix}0 & m+ i  wk \\ -m- i  wk & \end{pmatrix},
\end{equation}
where we recovered the parameter $w$ to compute the integral in the general case. Working out the trace in \cref{Eq: General Exp for Pol Winding} yields
\begin{align*}
    \nu&=\int_{-\infty}^\infty\frac{\dd k}{4\pi  i }\frac{2 i  mw}{m^2+w^2k^2} \\
    &=\int_{-\infty}^\infty\frac{\dd k}{2\pi }\frac{\mathcal{M}}{\mathcal{M}^2+k^2} \\
    &=\frac{\mathcal{M}}{2\pi}\int_{-\infty}^\infty \dd k\frac{1}{\mathcal{M}^2}\frac{1}{1+(k/\mathcal{M})^2},
\end{align*}
where we introduced the scaled mass parameter $\mathcal{M}=m/w$. Now we use the fact that $a/a^2=1/|a|$, and substitute $k=\mathcal{M}x$ to get
\begin{equation*}
    \nu=\frac{\mathcal{M}}{2\pi|\mathcal{M}|}\int_{-\infty}^\infty \dd x\frac{1}{1+x^2}=\frac{1}{2}\text{sign}(\mathcal{M}).
\end{equation*}
The last equality results from $a/|a|=\text{sign}(a)$ and the integral evaluating to $\pi$.

\section{Thirring and sine-Gordon models}
\label{App: Thirring and Sine-Gordon models}
In this section, we list some properties of the Thirring model. Most of them are obtained using its bosonic dual, the sine-Gordon model. Both are instances of integrable field theories, which are (1+1)-dimensional field theories that have an infinite number of local conserved charges.
Such operators commute with the respective Hamiltonian and will strongly constrain the scattering between particles and, in general, the dynamics in the model.
Scattering between particles is elastic and factorizes into subsequent two-particle scatterings.
These requirements are strict enough to analytically fix the form of the scattering matrix via bootstrap methods \cite{Mussardo2009Exact-Matrices}, which is why these theories are analytically solvable.
In particular, the Green's functions and the operator's matrix elements can be derived analytically.

Starting from the Lagrangians of the Thirring and sine-Gordon models (for simplicity, we consider $v_\mathrm{F} = 1$)
\begin{equation*}
    \begin{split}
        \mathcal{L}_\mathrm{Th} &= \bar{\Psi}(x)\left( i \slashed{\partial}-m_0\right)\Psi(x)-\frac{g}{2}\left[\bar{\Psi}(x)\gamma_\mu\Psi(x)\right]^2 , \\
        \mathcal{L}_\mathrm{SG} &= \frac{1}{16\pi}\left(\partial_\mu\Phi\right)^2 + \frac{m_0^2}{\beta^2}\left(\cos{\beta\Phi}-1\right), 
    \end{split}
\end{equation*}
the scattering matrix of Thirring fermionic excitations, labeled by $i=\pm$ when fermions ($+$) or antifermions ($-$) are considered, can be fully specified using inverse scattering methods. It is given by \cite{Korepin1979DirectModel,Korepin1993QuantumFunctions}
\begin{equation}
    \label{Eq: S-matrix Thirring}
    S_{ij}(\theta) = -e^{8 i /\beta^2}\frac{e^\theta-e^{-8 i /\beta^2}}{e^\theta-e^{8 i /\beta^2}}\delta_{ij} + (1-\delta_{ij}),
\end{equation}
where the argument $\theta$ is the difference between two rapidities $\theta_1$ and $\theta_2$, related to the momenta $p_1$ and $p_2$ through $p_i = m\cosh\theta_i$. The scattering matrix has the following action on a two-particle state
\begin{equation*}
    S_{i_1i_2}(\theta)|\theta_1,\theta_2\rangle_{i_1,i_2} = |\theta_2,\theta_1\rangle_{i_2,i_1}.
\end{equation*}
Moreover, given its action on states, we can infer that $S(-\theta) = S^\dagger(\theta) = S^{-1}(\theta)$, 
which holds for \cref{Eq: S-matrix Thirring}.
This expression depends also on the parameter $\beta^2$, which is just the coupling of the sine-Gordon model.
Moreover, we take the mass of the Thirring fermion $m$ to be a renormalized version of the mass parameter $m_0$, given by \cref{Eq: Physical Mass} and obtained by studying the sine-Gordon theory. Before turning to the latter, we briefly introduce how correlation functions are evaluated in the Thirring model: once the following matrix elements of a certain operator $V$ are known,
\begin{equation}
    \label{Eq: Form Factors}
    F^V_{n;i_1\dots i_n}(\theta_1,\dots,\theta_n) = \langle 0|V|\theta_1,\dots,\theta_n\rangle_{i_1\dots i_n},
\end{equation}
the Lehmann representation is fully described as an absolutely convergent series (after Wick rotating to Euclidean spacetime)
\begin{align}
    \label{Eq: Lehmann FF}
    G^V(x,t)& = \langle 0|V(x,t)V^\dagger(0,0)|0\rangle\\
    & = \begin{multlined}[t]
        \sum_{n=0}^\infty\int_{-\infty}^{+\infty}\dd\theta_1 \dots \dd\theta_n \\
        \times|F^V_n(\theta_1,\dots,\theta_n)|^2 e^{ i (kx+\Omega t)},
    \end{multlined} \notag
\end{align}
where $k = m\sum_{j=1}^n\sinh{\theta_j}$ and $\Omega = m\sum_{j=1}^n\cosh{\theta_j}$.
Because of their importance, the matrix elements of \cref{Eq: Form Factors} are dubbed form factors, and self-consistency arguments constrain their analytic form \cite{Smirnov1992FormTheory}.
One can also infer from the representation \cref{Eq: Lehmann FF} that two-point functions in momentum space do not contain zeros on the real axis, except for $k\to 0, \infty$.
Moreover, \cref{Eq: Lehmann FF} is crucial to verify scaling properties of Green's function in the UV and IR regions in \cref{Eq: Green's Function scaling massive}: the first can be studied using the Feynman gas method, and one can numerically check that predictions from CFTs are consistent with form factor computations, while the second is straightforward since, by performing a Wick rotation, one can show that every two-point function scales as $e^{-mL}$ in this regime, where $L$ is the system size. 

Now, to unveil the relations between the bare and the physical quantities of the Thirring model, we turn to the sine-Gordon model. Moreover, the bosonization procedure naturally lends itself to a more straightforward computation of the Green's function of the Thirring model, compared to the direct method.
We start by the bosonisation relations between the fermionic fields $\Psi$ and $\bar{\Psi}$ and the bosonic ones $\Phi$ and $\Theta$, where $\Theta$ is the dual field given by $-\partial_x\Theta(x,t) = \partial_t\Phi(x,t)$,

\begin{equation}\label{Eq: Bosonization}
    \begin{split}
            \Psi(x,t)&= 
    \exp{\frac{ i }{2}\left[\beta\Phi(x,t) + \frac{1}{\beta}\Theta(x,t)\right]},
   \\
   \bar{\Psi}(x,t) &= \exp{-\frac{ i }{2}\left[\beta\Phi(x,t) -\frac{1}{\beta}\Theta(x,t)\right]}
       .
    \end{split}
\end{equation}

In the main text, we employ then the following notation, borrowed from Ref.~\cite{Lukyanov2001FormModel}:
\begin{equation}\label{Eq: BosonizationLZ}
    \begin{split}
    \Psi(x,t)&=    e^{ i  \beta\Phi(x,t)/2}\mathcal{O}_{\beta/2}^1(x,t),\\
    \bar{\Psi}(x,t) &= e^{- i  \beta\Phi(x,t)/2}\mathcal{O}_{-\beta/2}^1(x,t) ,
    \end{split}
\end{equation}
where the operator $\mathcal{O}_{\pm B/2}^1$ contains an infinite product given by the Baker-Campbell-Hausdorff factorisation of the exponential in \cref{Eq: Bosonization}, and depends on the phase field $\Theta$ and all its commutators with $\Phi$.
Moreover, the origin of a string-like term in the fermionic creation/annihilation operators is given by the relation between the bosonic field and its dual,
\begin{equation}
    \label{Eq: Bos String}
    \Theta(x,t) = -\int_{-\infty}^x\dd s \partial_t\Phi(s,t).
\end{equation}
Such an operator appears in the exponent of \cref{Eq: Bosonization} and weights the configuration of the field from the point $x$ where it is placed, to $-\infty$.
We proceed finally by plugging in the definition \cref{Eq: Bosonization} into the Thirring Lagrangian in \cref{Eq: Thirring Lagrangian} to obtain
\begin{equation}
    \mathcal{L}_\mathrm{SG} = \frac{1}{2}\left[\left(\partial_x\Phi\right)^2 + \left(\partial_x\Theta\right)^2\right] + \frac{m_0^2}{\beta^2}\left(\cos{\beta\Phi}-1\right),
\end{equation}
Provided the following identity holds:
\begin{equation}
    \beta^2 = \frac{1}{2}\frac{1}{1-g/\pi},
\end{equation}
that allows us to relate the coupling constants from the two models.

The sine-Gordon model has another peculiarity, namely, it can be understood as the integrable deformation of a coset CFT $(\mathrm{G}_1\times\mathrm{G}_1)/\mathrm{G}_2$ \cite{Goodard1985VirasoroModels,Goddard1986UnitaryAlgebras}. This piece of information, joined with the study of the thermodynamics of the model using the thermodynamics Bethe Ansatz, allows one to compute the relation between mass and coupling constant.
Fateev developed the original argument for a broad class of models, which were refined by Zamolodchikov in the sine-Gordon case \cite{Fateev1994TheTheories}. Here, we briefly follow Mussardo \cite{Mussardo2009ThermodynamicalAnsatz} for a simple review of the main argument.
The main idea lies in computing the free energy of an integrable model both using the thermodynamics Bethe Ansatz and conformal perturbation theory (see \cref{App: Scaling dimension} for an introduction): in the first case, we can expand for small values of the mass $m$, while in the second we will expand in the coupling constant $g$.
We are then able to match the two series to obtain
\begin{equation*}
    \beta = \mathcal{D}m^{2-2\Delta},
\end{equation*}
which is expected from dimensional analysis (the conformal weight $\Delta$ belongs to the relevant deforming operator, i.e. $\cos{\beta\Phi}$ in the sine-Gordon case and $\Delta_\mathrm{SG} = \beta^2$), but the meaningful prefactor $\mathcal{D}$ can be directly computed.
We stress that such a result cannot be obtained directly from the Thirring model itself and proves once more the power of bosonization methods.

\section{Derivation of the scaling behavior of Green's function in momentum space}
\label{App: Derivation}
We start by considering the Fourier transform of the Green's function, i.e., Green's function in momentum space, 
\begin{equation*}
    G(\mathbf{k}) = \int_{\mathbb{R}^{2}}\frac{\dd \mathbf{k}}{(2\pi)^2}G(\mathbf{x})e^{- i \mathbf{k}\cdot\mathbf{x}},
\end{equation*}
where spacetime is equipped with a Euclidean signature. We now split the integration region into three different ones:
\begin{enumerate}
    \item The UV region $|x|<a$,
    \item The regular region $a<|x|<R$,
    \item The IR region $|x|>R$.
\end{enumerate}
From \cref{Eq: Green's Function scaling massive}, we know the behavior in the scaling regions, and by redefining the integration variables $\mathbf{k}\cdot\mathbf{x}=kr\cos\theta$, with $kr=s$, we obtain in the UV region
\begin{equation*}
    G_\mathrm{UV}(\mathbf{k}) = k^{y_\mathrm{uv}-2}\int_{0}^{+ka}\frac{\dd s}{2\pi}\int_0^{2\pi}\frac{\dd\theta}{2\pi}s^{y_\mathrm{uv}-1}e^{- i  s\cos\theta},
\end{equation*}
where $y_\mathrm{uv}$ is the scaling dimension in the UV regime. The only momentum dependence left in the integral is in the extremes of integration.
When the system approaches the critical phase, it is only determined by its UV completion. As such, we can consider the limit $a\to\infty$, and the only dependence is power-law-like (as we would be expecting from a simple scaling argument \cite{Cardy1996ScalingPhysics}).
For the same reason, this expression is also a pure power-law when the momentum is large.

When the IR region is considered, the same considerations occur, except for the presence of different integrands depending on whether we consider massive or massless theories.
For a massive theory, the result would read
\begin{equation*}
    G_\mathrm{IR}(k) = e^{-mR}\left(\frac{e^{ i  kR}}{k+ i  m}-\frac{e^{- i  kR}}{k- i  m}\right),
\end{equation*}
and the region for $k\to0$ is fully described by the constant result proportional to $1/m$.
For gapless systems, this result is just
\begin{equation*}
    G_\mathrm{IR}(k) = k^{y_\mathrm{ir}-2}\int_{kR}^{+\infty}\frac{\dd s}{2\pi}s^{y_\mathrm{ir}-1}e^{-is},
\end{equation*}
and it ensures again a power-law behavior. For pure CFTs, $y_\mathrm{uv}=y_\mathrm{ir}$.

Our treatment does not affect the regular region; the only requirement that it must fulfill is the absence of zeros or singularities, which is ensured in integrable models.

\section{Scaling dimension in quantum field theory}
\label{App: Scaling dimension}
An unexpected outcome of the renormalization procedure in quantum field theories is the presence of anomalous dimensions of local fields.
This effect can be understood from two different points of view: at the level of a lattice, in which the rescaling of the lattice spacing induces a change in the theory's parameters, or at the level of the field theory, where the couplings run under rescaling of the momentum space.

Even though the relation between the two is, at first sight, straightforward (the lattice spacing and the momentum cutoff of a theory are reciprocal to each other), it is highly relevant to the understanding of our results: the invariance from lattice spacing and momentum cutoff informs us about the necessity of describing our topological invariant/polarization only in terms of scale-invariant quantities, i.e. quantities that are defined in scaling invariant regions of the theory. In addition, the realization of the quantum field theory on a lattice and the insertion of a momentum cutoff are both instances of a regularization procedure needed when interactions are relevant in a quantum field theory. We shall now outline why and how scale-invariant regimes are present in lattice and continuum theories \cite{Collins1984BasicExamples}.

\paragraph{Lattice.} Consider a lattice of length $L=Na$, where $N$ is the number of sites and $a$ is the lattice spacing. Imposing that the universal properties of the physical system contained in its partition function are invariant under rescaling, we can perform a change of the lattice spacing $a\to a' = ba$ and the number of sites $N\to N'$.
This procedure is called a real-space renormalization group. To keep the partition function of the system unchanged, all the parameters describing the Hamiltonian of the theory will change accordingly in a generally complicated and nonlinear transformation $g_i\to g_i' = f_i(\{g_k\},b)$.
The iteration of this procedure leads to a family of points in the coupling constant manifold representing the same system, observed with different values of the parameters $a$ and $N$. 
Since the correlation length of the system is simply rescaled,
\begin{equation*}
    \xi(g') = b^{-1}\xi(g),
\end{equation*}
the fixed points of the RG, $g_i^* = f_i(\{g_k^*\},b)$, imply that the correlation length vanishes (trivial points) or diverges (critical points) across them.
This, in turn, implies that the physical dimension of operators on the lattice changes along the renormalization group flow. An operator on the trivial point has a dimension equal to its physical dimension, established by means of usual dimensional counting, while one at the critical dimension is determined by critical exponents. These two values are usually different, and this difference is accounted for by the anomalous dimension.
The reason for this phenomenon is better understood in continuum theories. \\

\paragraph{Continuum.} In a field theory, performing the renormalization procedure in momentum space is simpler.
Here, divergences in Feynman diagrams must be accounted for by introducing counterterms in the Lagrangian, resulting in a shift of the couplings and leading to similar conclusions.
These divergences in Feynman diagrams parts -- usually at high momenta -- require a regularization procedure, typically achieved by imposing a momentum cutoff $\Lambda$. Following this procedure, the sources of singular behavior can be safely removed at the price of introducing a dependence between the couplings -- and the fields -- on the cutoff via counterterms, i.e., $g_i\to g_i'=f_i(\{g_k\},\Lambda)$.
Adding new counterterms at higher orders of the coupling constants is equivalent to shifting the cutoff value toward infinity (which is the point where it is formally removed).
On the other hand, we can also squeeze our momentum space by rescaling the cutoff $\Lambda\to\Lambda'=\Lambda/b$; we would get the same picture of the lattice procedure, and we could study the relation between couplings and cutoff in detail without the need to compute more counterterms.
Expanding the knowledge of the UV-region $k\in(\Lambda,+\infty)$ with this procedure allows studying the properties of the model close to the critical fixed point; it is then simple to draw the picture we used in \cref{App: Derivation} of the whole momentum space split into regular part (determined by the massive interacting theory), IR-region (ruled by the trivial fixed point as a massive free theory) and the UV-region (described by the critical fixed point as a massless interacting theory).
Moreover, as elucidated in the main text, imposing the invariance of expectation values (namely correlation functions) leads to a set of equations [see \cref{Eq: CSequation}] that rule the renormalization group flow in the momentum space; among these, the presence of counterterms for fields induces a non-trivial anomalous dimension, especially at the fixed point where $f_i(g^*)=0$.
Now, we can understand the presence of an anomalous dimension as the effect of mixing between the original free field with others needed to remove singularities arising from turning on the interaction; in other words, the dressing of the free field induces the anomalous dimension.\\

It is then clear why taking the continuum limit of the interacting SSH model would not spoil its topology. Since the polarization is a scale-invariant quantity, it is customary to consider the limit for $a\to 0$, where the Thirring model is retrieved. More than that, the polarization must retain scale-independent properties from fixed points related to the Thirring model, i.e., its massive free fermionic theory in the IR regime and the massless Thirring model at the UV point.

In addition, one can check the scaling properties of Green's functions along the RG trajectory that connects the two fixed points, as in \cref{Eq: Green's Function scaling massive}, where the scaling dimension of fermionic creation and annihilation operators enters. The form factor expansion fully controls the IR point, so we just check the UV, which is the one depending on the anomalous dimension.
The idea lies in considering the fixed point (which, for simplicity, we consider described by a CFT and not just critical; this is the case for Thirring and the sine-Gordon model) and a perturbation that drives the system away from it by opening a gap in the spectrum.
Note that the above-mentioned integrable theories can also be expressed close to their critical fixed points as gap-opening perturbations of conformal field theories.
CFTs are fully solvable, and the perturbation we add,
\begin{equation*}
    \mathcal{L} = \mathcal{L}_\mathrm{CFT} + \lambda V(r) ,
\end{equation*}
can be treated perturbatively for arbitrary small values of $\lambda$. Hence, one can reconstruct exactly the terms of the series for expectation values of local operators $A_k$:
\begin{equation*}
    \langle A_k\rangle_\lambda = \frac{1}{Z_\lambda}\sum_{n=0}^\infty\frac{(-\lambda)^n}{n!}\int\dd r_1\dots\dd r_n 
    \langle A_k V(r_1)\dots V(r_n)\rangle_0 .
\end{equation*}
Then, we can extract the most diverging term of the Green's function in the limit $r\to 0$, studying the operator product expansion in the off-critical region,
\begin{equation*}
    \Psi(z,\bar{z})\bar{\Psi}(0) = \sum_k C_{\Psi\bar{\Psi}}^k(z,\bar{z}) A_k(0) .
\end{equation*}
We then obtain, based on dimensional analysis,
\begin{equation*}
    C_{\Psi\bar{\Psi}}^k(z,\bar{z}) = z^{\gamma_k-2\gamma_\Psi} \bar{z}^{\gamma_k-2\bar{\gamma}_\Psi}
    \sum_{n=0}^\infty D_{\Psi\bar{\Psi}}^{k,n}(\lambda r^{2-\gamma_\Psi - \bar{\gamma}_\Psi})^n ,
\end{equation*}
where $r=(z\bar{z})^{1/2}$.
Then, the most diverging term occurs for $n=0$ and $\gamma_k = 0$, which involves the identity operator $A_0 = \mathbf{1}$.
We have shown that in the limit $r\to0$, the dominant contribution is still given by a power-law determined by the CFT that describes the fixed point.
Such a statement is valid at any level of perturbation theory and directly leads to \cref{Eq: Green's Function scaling massive}.

\end{document}